\documentclass{INTERSPEECH2023}

\newif\ifarxiv
\arxivfalse

\interspeechcameraready

\usepackage{color}
\usepackage{comment}
\usepackage{multirow}
\usepackage{makecell}
\usepackage{subfig}
\usepackage{soul}
\usepackage{mathtools}
\usepackage{bigstrut,multirow,rotating}
\usepackage{algorithm} 
\usepackage[noend]{algpseudocode} 

\newtheorem{definition}{Definition}

\title{Efficiently Train ASR Models\\ that Memorize Less and Perform Better with Per-core Clipping}
\name{Lun Wang, Om Thakkar, Zhong Meng, Nicole Rafidi, Rohit Prabhavalkar, Arun Narayanan}
\address{Google}
\email{
\{lunwang, omthkkr, zhongmeng, nrafidi, prabhavalkar, arunnt\}@google.com
}

\begin{document}

\maketitle
 
\begin{abstract}
Gradient clipping plays a vital role in training large-scale automatic speech recognition (ASR) models. It is typically applied to minibatch gradients to prevent gradient explosion, and to the individual sample gradients to mitigate unintended memorization. This work systematically investigates the impact of a specific granularity of gradient clipping, namely per-core clipping (PCC), across training a wide range of ASR models. We empirically demonstrate that PCC can effectively mitigate unintended memorization in ASR models. Surprisingly, we find that PCC positively influences ASR performance metrics, leading to improved convergence rates and reduced word error rates. To avoid tuning the additional hyperparameter introduced by PCC, we further propose a novel variant, adaptive per-core clipping (APCC), for streamlined optimization. Our findings highlight the multifaceted benefits of PCC as a strategy for robust, privacy-forward ASR model training.
\end{abstract}


\section{Introduction}
\label{sec:intro}

While large neural networks start to exhibit emergent abilities, many generative vision and language models have been found to regurgitate their training data during inference~\cite{carlini2019secret,carlini2021extracting,carlini2023extracting,nasr2023scalable}.
Therefore, preserving user privacy within the training data of these models has become a critical and widely-studied concern recently.
For non-generative automatic speech recognition (ASR) models, there are also a few recent works demonstrating corresponding privacy attacks~\cite{amid2022extracting,jagielski2024noise,wang2024unintended}, indicating that speech models are not spared from such privacy risks.

The gold standard of privacy for neural networks is differential privacy (DP)~\cite{dwork2006calibrating}, and the workhorse for differentially private large-scale deep learning is differentially private stochastic gradient descent (\emph{i.e.} DP-SGD)~\cite{abadi2016deep}.
However, the strong privacy of DP-SGD comes to the detriment of computational cost~\cite{li2021large} and utility overhead~\cite{abadi2016deep}, which can be prohibitive in many potential use cases such as training large-scale ASR models.
To address the issue, a recent line of work~\cite{huang2022detecting,wang2024unintended} proposes to omit the noise addition operation in DP-SGD and only keep the per-example clipping operation for better utility at the cost of providing only empirical privacy.
Surprisingly, per-example clipping on its own is able to significantly mitigate state-of-the-art attacks on ASR models~\cite{huang2022detecting,wang2024unintended}.

However, per-example clipping still suffers from computational overhead.
The reason is that to conduct per-example clipping, the gradients of all training examples, namely \textbf{per-example gradients}, need to be materialized in memory.
However, most common deep learning frameworks avoid completely materializing per-example gradients for speed and memory optimization~\cite{subramani2021enabling} because such materialization places a substantially higher demand on compute resources.
As the model size grows larger, such overhead gradually becomes unsustainable.
Although there has been a line of work attempting to reduce the memory/compute overhead of per-example clipping, the solutions either incur other trade-offs, such as an extra back-propagation pass~\cite{li2021large,bu2022scalable} or worse utility~\cite{bu2021fast,he2022exploring}, or do not completely alleviate the overhead~\cite{subramani2021enabling}.

To bypass the issue, we make a key observation that almost all large ASR models are trained with data parallelism~\cite{chong2010opportunities}.
In data parallelism, a mini-batch of training examples is sharded across a few compute cores (\emph{e.g.} GPUs/TPUs).
Each compute core runs forward- and backward-propagation on its own data shard to compute the average gradient of the shard, and then all the compute cores  synchronize to obtain the average mini-batch gradient.
This implies, even in non-private training setups, each core needs to materialize its own average gradient per training step, namely per-core gradient, before aggregating across cores.
Therefore, if we clip the per-core gradients instead of per-example gradients, it only incurs negligible memory/compute overhead.

The idea of per-core clipping (PCC) can be viewed as a special case of micro-batch clipping~\cite{mcmahan2018general, Ponomareva_2023} in model training with data parallelism.
Traditional micro-batch clipping is also designed to reduce the compute/memory overhead of DP-SGD at the expense of a lower signal-to-noise ratio.
PCC, by leveraging the fact that per-core gradients always get materialized in training with data parallelism, almost completely eliminates the compute/memory overhead.

However, a surprising observation from our experiments is that \textbf{PCC can improve the performance of ASR models}.
Concretely, we conducted an extensive evaluation of PCC across several ASR models with different architectures and datasets, and observed faster convergence rate and lower WER in most of them.
Our conjecture for the improvement is that per-core clipping acts as an implicit regularization that prevents the extreme outliers from slowing down the convergence or deviating the model from a better local minima.
We leave a thorough investigation into the factors contributing to the observed improvement to future work.

PCC introduces one hyperparameter, the clipping bound, which can require tuning for best results.
To address this, we propose a hyperparameter-free variant of per-core clipping, namely adaptive per-core clipping (APCC).
By using the minimum L2 norm across the per-core gradients in each iteration as the clipping bound, APCC removes the extra hyperparameter and even further improves the performance metrics.

\begin{algorithm*}
	\caption{Pseudocode for minibatch-SGD with PCC/APCC and data parallelism (changes for PCC in \textcolor{blue}{blue}, and APCC in \textcolor{red}{red}). $\bar{g}$ represents per-core gradients. $\hat{g}$ represents clipped per-core gradients. $g$ represents mini-batch gradients.}
	\label{alg:pcc}
	\textbf{Input: initial parameters $w_0$, loss function $\mathcal{L}$, training data $\mathcal{D}$, \#iterations $T$, per-core batch size $B$, \#compute cores $C$, learning rate $r$, \textcolor{blue}{clipping bound $b$}.}
	\begin{algorithmic}[1]
		\For {$t=1,2,\ldots T$}
		    \State $\{d^t_j\}_{j \in \{1, \ldots, BC\}}\leftarrow\mathcal{D}$\Comment{sample mini-batch from $\mathcal{D}$}
		    \For {$c=1,2,\ldots C$ \textbf{parallelly}}
		         \State The $c^{th}$ core loads its data shard $\{d^t_j\}_{j \in \{B(c-1)+1, \ldots, Bc\}}$
		         \State $\bar{g}_t^c = \nabla \mathcal{L}(w_{t-1}, \{d^t_j\}_{j \in \{B(c-1)+1, \ldots, Bc\}})$\Comment{forward/backward propagation on each compute core}
		         \textcolor{blue}{\State $\hat{g}_t^c = \frac{\text{min}(||\bar{g}_t^c||_2, b)}{||\bar{g}_t^c||_2}\cdot\bar{g}_t^c$\Comment{per-core clipping}}
		    \EndFor
		    \textcolor{red}{\State $b_t = \min_{c\in\{1,\ldots,C\}}||\bar{g}_t^c||$
		    \For {$c=1,2,\ldots C$ \textbf{parallelly}}
		         \State $\hat{g}_t^c = \frac{ b_t}{||\bar{g}_t^c||_2}\cdot\bar{g}_t^c$\Comment{adaptive per-core clipping}
		    \EndFor}
			\State $g_t = \sum_{c\in\{1,\ldots,C\}}\hat{g}_t^c$\Comment{cross-core aggregation}
			\State $w_t = w_{t-1} - r\cdot g_t$\Comment{update the model parameters}
		\EndFor
	\end{algorithmic} 
\end{algorithm*}

In summary, we make the following contributions:
\begin{itemize}
    \item We propose per-core clipping, a variant of gradient clipping that suppresses unintended memorization in ASR models with negligible compute overhead. We showcase that per-core clipping can effectively mitigate unintended memorization on a Conformer~\cite{gulati2020conformer} model fine-tuned on LibriSpeech~\cite{panayotov2015librispeech}.
    \item We perform a comprehensive empirical assessment of PCC across a diverse set models, and observe performance and convergence rate enhancements.
    \item We propose adaptive per-core clipping, a variant of PCC releasing the burden of extra hyperparameter-tuning.
\end{itemize}
\begin{table*}[th]
    \caption{Exposure of canaries by CER with different insertion frequencies. Mean and standard deviation are calculated from 20 canaries. \textbf{Bold} highlights best.}
    \label{tab:exposure}
    \centering
    \begin{tabular}{c|c|c|c|c|c}
    \toprule
    Canary \#Insertion & 1 & 2 & 4 & 8 & 16 \\
    \midrule
    Baseline & $4.8\pm3.7$ & $11.0\pm3.8$ & $13.0\pm3.5$ & $13.2\pm3.4$ & $13.5\pm2.9$ \\
    PCC@2.5 & $\textbf{1.0}\pm0.0$  & $\textbf{1.0}\pm0.0$ & $\textbf{1.0}\pm0.0$ & $\textbf{1.5}\pm2.3$ & $\textbf{2.1}\pm3.2$ \\
    APCC & $1.7\pm1.8$ & $1.7\pm1.3$ & $1.3\pm1.5$ & $2.2\pm1.5$ & $2.5\pm2.3$\\
    \bottomrule
    \end{tabular}
\end{table*}

\begin{table}[th]
  \vspace{-5pt}
  \caption{Best WER on LibriSpeech within 20K fine-tuning steps. Mean and standard deviation are calculated across 3 runs. \textbf{Bold} highlights best.}
  \label{tab:publicwer}
  \centering
  \begin{tabular}{c|c|c|c}
    \toprule
    Test set & Baseline & PCC & APCC \\
    \midrule
    test-clean & 1.93 $\pm$ 0.04 & 1.87 $\pm$ 0.05 & \textbf{1.85} $\pm$ 0.07 \\
    test-other & 3.99 $\pm$ 0.01 & \textbf{3.80} $\pm$ 0.04 & \textbf{3.80} $\pm$ 0.05 \\
    \bottomrule
  \end{tabular}
  \vspace{-5pt}
\end{table}

\section{Mitigating Memorization via PCC}
\label{sec:privacy}

In this section, we motivate and describe the design of PCC, and show its empirical privacy advantage under a SOTA attack~\cite{wang2024unintended}.

\subsection{Reducing DP-SGD's Overheads for Empirical Privacy}
As mentioned in Section~\ref{sec:intro}, DP-SGD~\cite{abadi2016deep} has been a workhorse for large-scale differentially private deep learning since its proposal in 2016.
Compared to other DP deep learning algorithms such as DP-FTRL~\cite{kairouz2021practical} and DP-MF~\cite{denisov2022improved,choquette2022multi,choquette2023amplified} which can achieve better privacy-utility trade-offs by using correlated noise across iterations, the main advantage of DP-SGD is that it is stateless, resulting in a smaller compute overhead.

Compared to its non-private counterpart, DP-SGD still implicitly incurs conspicuous compute/memory and utility overheads.
The utility overhead mainly stems from the random noise addition in DP-SGD.~\cite{bst14}
One solution to this is to remove the noise addition operation, and only keep the per-example clipping (PEC) operation in DP-SGD. While such training no longer satisfies any meaningful DP, a recent line of work~\cite{carlini2019secret,huang2022detecting,wang2024unintended} has shown that
models trained in this manner tend to match the utility of baseline models while being
empirically less prone to memorization.

However, the PEC approach still incurs a compute overhead.~\cite{subramani2021enabling} The PEC operation requires materializing per-example gradients, which adds a high memory overhead on the training devices.
Although there have been a few techniques~\cite{li2021large,bu2021fast,bu2022scalable,he2022exploring} proposed to avoid/reduce the overhead, they incur other trade-offs, such as an extra round of back-propagation, or worse utility.

\subsection{Per-core Clipping}

To overcome the compute overhead of PEC, we make a key observation that nowadays, many large ASR models are trained with data parallelism~\cite{chong2010opportunities}.
Consider such a setting (shown in
~Algorithm~\ref{alg:pcc}) where, for each training step, a mini-batch of training examples are sharded on multiple compute cores.
To obtain a mini-batch gradient, aggregated gradients (called \emph{per-core gradients}) are computed at each core before undergoing a cross-core aggregation. 
Thus, an operation of Per-core Clipping (PCC), i.e., clipping the per-core gradients, does not incur any memory overhead to the training pipeline.

While the memory advantage of PCC over PEC increases with the number of training examples in each core (\emph{i.e.,} the per-core batch size), the effect of mitigating memorization can reduce as a result.
Think about the two extreme cases: 1) When per-core batch size is 1, PCC reduces to PEC and provides exactly the same trade-offs; and 2) when per-core batch size equals the mini-batch size (i.e., no data parallelism), PCC  reduces to purely a learning rate scaling operation without affecting the minibatch gradients.

\subsection{Measuring Unintended Memorization} \label{sec:exposure}

\noindent\textbf{Secret Sharer \& Exposure:}
To study the empirical privacy benefits of PCC, we use the Secret Sharer framework~\cite{carlini2019secret}, and follow the methodology specific to ASR models in \cite{wang2024unintended}.
Concretely, we use a WaveNet Text-to-Speech (TTS) engine~\cite{oord2018parallel} to generate 4x-sped-up utterances, and insert them into the training set of the ASR models.
The transcripts fed to the TTS engine are a combination of random words of length 7 following Wang \emph{et al.}~\cite{wang2024unintended}.
These utterances as called canaries, and the goal is to design them to be out-of-distribution enough from the regular training examples so that a model is not able to generalize to them without seeing them during training.
If an ASR model demonstrates unexpectedly high accuracy in transcribing canary utterances, this suggests strong evidence that the model has engaged in unintended memorization of training data rather than developing robust generalization capabilities.
Moreover, such memorization can make the model susceptible to leakage about its training data via privacy attacks.

In this work, we measure the trained ASR model's character error rate (CER) of both the canaries and utterances drawn from the same distribution but unseen during training (\emph{i.e.,} a heldout set).
If the model does exceptionally well on the canaries compared to heldout set, this indicates the model has memorized the canaries.
This intuition can be formally captured by exposure, a metric used to measure such memorization.
The higher the exposure of a canary, the stronger it has been memorization. 

\begin{definition}[Exposure~\cite{carlini2019secret}]
Given a canary $c$, a model $\mathcal{M}$, and examples in a holdout set $r_i$, the exposure of $c$ is
\begin{equation*}
\textbf{exposure}_\mathcal{M}(c, \{r_i\}) = \log_2|\{r_i\}| - \log_2rank_\mathcal{M}(c, \{r_i\}),
\end{equation*}
 where $|\{r_i\}|$ is the size of the holdout set, and $rank_\mathcal{M}(c, \{r_i\})$ is the rank of canary $c$ among $r_i$ in terms of a metric of interest, e.g., loss, or CER.
\end{definition}

\noindent\textbf{Experimental Setup:}
We use the 600M Conformer XL~\cite{zhang2020pushing}, a state-of-the-art ASR model architecture, for our memorization analysis.
The encoder is pre-trained on the LibriLight dataset~\cite{kahn2020libri} for 1 million steps using the BEST-RQ~\cite{chiu2022self} self-supervised training technique.
Next, we attach a decoder to the model, and fine-tune the complete model on the LibriSpeech dataset~\cite{panayotov2015librispeech} for 20,000 steps.
The canaries are inserted during the fine-tuning phase, and the training approach mirrors~\cite{wang2024unintended}.
The per-core batch size used is 4, and 128 cores are used to train each model.

\noindent\textbf{Evaluation Results:}
We train Conformer XL with and without PCC, and the first 2 rows in Table~\ref{tab:exposure} summarize the privacy evaluation results.
The clipping bound 2.5 is selected based on WER on LibriSpeech test-other by grid searching $\{1, 2.5, 5, 10, 100\}$ on Conformer M~\cite{gulati2020conformer}, a 20x smaller model with a similar architecture.
We use 2.5 for all the following experiments without further tuning the hyper-parameter.
We first observe that for baseline training, even canaries appearing only once in the training set have an average exposure of 4.8, while for per-core clipping, exposure is at its lower bound (i.e., no detection of unintended memorization) until 4 repetitions of the canary in the training set.
We also notice that the standard deviation of exposure for the model trained with PCC is 0.0 for canaries appearing at most 4 times in the training set.
To understand this better, we manually inspected the decoding output for such canaries by the PCC model, and found that the model outputs empty for the inserted fast canaries, perhaps an ideal behavior when encountering indecipherable utterances.

\subsection{Adaptive Per-core Clipping}
Training ASR models can encompass a variety of architectures and datasets, and hence lead to different gradient norms during training.
As a result, deploying PCC can involve tuning  the clipping bound.
To alleviate such burden, we devise a variant called Adaptive PCC (APCC), which adaptively uses the minimum L2-norm among all the per-core gradients as the clipping bound for each training step, as shown in details in Algorithm~\ref{alg:pcc}.

We ran the same exposure analysis as in Section~\ref{sec:exposure} for APCC, and the results are summarized in the 3rd row of~Table~\ref{tab:exposure}.
We observe that APCC also shows much lower exposure compared to the baseline model.
However, the exposure numbers are higher than PCC with clipping bound 2.5.
We conjecture this behavior could be attributed to the fact that the minimum per-core gradient norm is used without any privacy protection. However, we leave this investigation for future work.
\begin{table*}[th]
  \caption{Conformer for Voice Search Benchmark w/ or w/o PCC. \textbf{Bold} highlights better.}
  \label{tab:usm4vs}
  \centering
  \begin{tabular}{lcccccc|cccccc}
    \toprule
    Model & \multicolumn{6}{c}{Baseline} & \multicolumn{6}{c}{PCC} \\
    \midrule
    & VS & RM & RN & RP & RQ & RY & VS & RM & RN & RP & RQ & RY \\
    \midrule
    A0 & 4.0 & 13.0 & 14.8 & 37.7 & 20.8 & 23.9 & \textbf{3.8} & \textbf{12.7} & \textbf{12.8} & \textbf{37.2} & \textbf{20.5} & \textbf{23.4} \\
    A1 & 3.8 & 14.2 & 15.6 & 37.0 & 19.9 & 23.2 & \textbf{3.7} & \textbf{12.2} & \textbf{15.5} & \textbf{36.7} & \textbf{19.4} & \textbf{22.9} \\
    A2 & 3.7 & 10.3 & 16.1 & 33.1 & 16.2 & \textbf{20.1} & \textbf{3.6} & \textbf{9.9} & \textbf{14.1} & \textbf{31.5} & \textbf{15.1} & 20.3 \\
    \bottomrule
  \end{tabular}
\end{table*}

\begin{table*}[th]
  \caption{Modular Domain Adaptation for Voice Search Benchmark w/ or w/o PCC. \textbf{Bold} highlights better.}
  \label{tab:adapter4vs}
  \centering
  \begin{tabular}{lcccccc|cccccc}
    \toprule
    Model & \multicolumn{6}{c}{Baseline} & \multicolumn{6}{c}{PCC} \\
    \midrule
    & VS & RM & RN & RP & RQ & RY & VS & RM & RN & RP & RQ & RY \\
    \midrule
    MDA (en-us) & 5.2 & 4.1 & 15.9 & 11.3 & 25.0 & 26.6 & \textbf{5.1} & \textbf{4.0} & \textbf{15.8} & \textbf{11.1} & \textbf{24.9} & \textbf{26.5} \\
    MDA (fr-fr) & 9.5 &-&-&-&-&-& \textbf{9.2} &-&-&-&-&- \\
    \bottomrule
  \end{tabular}
\end{table*}

\section{Better ASR performance via PCC}
\label{sec:utility}

Since PCC provides better empirical privacy with no extra compute overhead, a natural question to ask is whether it causes any regression in ASR performance like DP-SGD.
In this section, we evaluate PCC across a variety of model architectures and datasets.
To our surprise, we find that PCC consistently improves their ASR performance. 

\subsection{Case Study 1: Conformer on LibriSpeech}

We first evaluate the model used in Section~\ref{sec:privacy}, the 600M Conformer XL, and provide the results in Table~\ref{tab:publicwer}.
We observe that on LibriSpeech test-other, after adding PCC, the average WER is improved from 3.99 to 3.80, a 4.8\% relative improvement.
APCC achieves the same average WER of 3.80.
On LibriSpeech test-clean, the WER is also improved using PCC/APCC by 3.1\%/4.1\% relative, respectively. 
Note that the standard deviations are small for all settings.

\subsection{Case Study 2: E2E ASR Model for Voice Search}\label{sec:e2e_vs}
For our second case study, we evaluate the effectiveness of per-core clipping on a large-scale voice search task.

\noindent\textbf{Model Architecture:}
The models for this case study follow the architecture described in Prabhavalkar \emph{et al.}~\cite{rohit2024extreme}.
Concretely, the models are hybrid autoregressive transducers (HAT)~\cite{variani2020hybrid}, comprised of an encoder, a prediction network, and a joint network.
The encoder is comprised of a convolutional sub-sampling block, followed by a series of 16 conformer blocks whose order of the convolution and multi-headed self-attention layers are exchanged.
The prediction network is a $V^2$ embedding prediction network~\cite{botros2021tied}.
The joint network is a linear layer followed by tanh activation.
More details about the model architecture can be found in~\cite{rohit2024extreme}.

\noindent\textbf{Training and Evaluation Sets:}
The training set is en-us utterances extracted from Google voice search traffic.
%
The majority of utterances are pseudo-labeled using a teacher model~\cite{hwang2022pseudo}, while a small portion is anonymized and human-transcribed, following Google AI principles~\cite{google_ai_principles}.
The test set is composed of 6 datasets, a voice search test set (VS), corresponding to the ``head" of the utterance distribution, and 5 rare word test sets containing rare words from different domains including maps (RM), news (RN), Google Play (RP), search queries (RQ), and YouTube (RY), corresponding to the ``tail" of the utterance distribution.
The rare word test sets are generated following the recipe described in~\cite{peyser2020improving}.

\noindent\textbf{Evaluation Results:}
We train the described model with and without PCC.
The evaluation results are summarized in Table~\ref{tab:usm4vs}, where model A0 corresponds to B0 in~\cite{rohit2024extreme}, A1 corresponds to E7~\cite{rohit2024extreme}, and A2 is A1 augmented with text injection~\cite{peyser2023improving}.
512 compute cores are used to train the model and the per-core batch size is 8 for A0 and A1, and 16 for A2.
We can observe that across all the architectures, the models trained with PCC almost always achieve a better WER on both voice search set and the rare word sets.
On A0, PCC achieves 5\%/4.1\% average relative improvement in WER on VS/rare word test sets.
On A1, PCC achieves 2.6\%/3.9\% average relative improvement in WER on VS/rare word test sets.
On A2, PCC achieves 2.7\%/5.4\% average relative improvement in WER on VS/rare word test sets.

\subsection{Case Study 3: Modular Domain Adaptation for Streaming Voice Search}

Now, we evaluate on the voice search task described in Section~\ref{sec:e2e_vs} above, but using the model architecture following ~\cite{li2023modular}.

\noindent\textbf{Model Architecture:}
We train streaming Conformer transducers~\cite{gulati2020conformer} following the recipe in~\cite{li2023modular}.
The encoder of the transducer consists of 7 causal Conformer blocks followed by 10 non-causal Conformer blocks~\cite{narayanan2021cascaded,sainath2022improving}.
%
%
%
%
There are two separate hybrid autoregressive transducer (HAT) decoders~\cite{variani2020hybrid} for the causal and non-causal encoders.
%
%
%
%
%
%
More details about the model architecture can be found in~\cite{rohit2024extreme}.

\noindent\textbf{Training and Evaluation Sets:}
The model is first trained on YouTube (YT) data to obtain a backbone model. 
%
%
Then, the backbone model is fine-tuned on voice search (VS) dataset as described in Case Study 2.
\ifarxiv
\else
Speaker tags~\cite{arumugam2023improved} were used during training to improve performance.
\fi
Note that we have two languages in Table~\ref{tab:adapter4vs}.
When training the en-us (fr-fr) model, both YT and VS dataset are mainly composed of en-us (fr-fr) utterances.
We do not have fr-fr rare word test sets.

\noindent\textbf{Evaluation Results:}
We train the described model with and without PCC.
128 cores are used to train each model, and the per-core batch size is 32.
We can observe from the results in Table~\ref{tab:adapter4vs} that after adding PCC, the WER is consistently improved across both languages and all the test sets.
The en-us model's performance on VS gets a 2.1\% relative improvement, and a 1.1\% average relative improvement on the rare word datasets.
The fr-fr model's performance on VS sees a 3.2\% relative improvement.

\subsection{Discussion: ASR Improvement by PCC}

While PCC consistently improves ASR quality, the reasons behind this improvement remain open for investigation.
We hypothesize that PCC reduces the impact of abnormally large gradients typically caused by extreme outliers (e.g., noisy utterances, foreign language content, music).
A crucial question is whether this improvement comes at the cost of performance on tail examples.
However, our findings in Case Studies 1 \& 2 suggest PCC does not sacrifice performance on tail examples.
Rather, we observe enhanced performance on rare word test sets.
This suggests a regularization-like dynamic, where PCC suppresses overfitting to extreme outliers in the training set while improving generalization to unseen tail examples.
To verify this, we measured WER on the training set~\cite{panayotov2015librispeech} and generalization error (\emph{i.e.} the difference of WER between test and training set) of the models outlined in Section~\ref{sec:exposure}.
The results in Table~\ref{tab:gen_gap} support the hypothesis: PCC exhibits typical regularization behavior: higher training loss and smaller generalization error.
This indicates PCC acts as an implicit regularization method.
We leave investigation into the connection between PCC and traditional regularization methods to future work.


\begin{table}[ht]
    \vspace{-5pt}
    \centering
    \caption{Generalization Gap w/ or w/o PCC.}
    \label{tab:gen_gap}
    \begin{tabular}{c|c|c}
    \toprule
    & WER on train-clean & Generalization Gap \\
    \midrule
    Baseline & $1.17\pm0.18$ & $3.00\pm0.16$ \\
    PCC@2.5 & $1.34\pm0.18$ & $2.63\pm0.15$ \\
    \bottomrule
    \end{tabular}
    \vspace{-5pt}
\end{table}
\section{Conclusion \& Future Directions}

In this work, we explored the challenges associated with mitigating unintended memorization and maintaining computational efficiency during the training of large-scale ASR models.
We proposed PCC, which can effectively mitigate such memorization with negligible computational overhead.
Surprisingly, we also observed PCC to improve the performance of a variety of ASR models, lowering WER.
We also introduced APCC to alleviate extra hyperparameter tuning.

\bibliographystyle{IEEEtran}
\bibliography{mybib}

\end{document}